\documentstyle[aps,twocolumn,epsf,floats,graphicx]{revtex}
\tightenlines

\def\order{{\cal O}}

\begin{document}
\title{Feedback Effect on Landau-Zener-St\"uckelberg Transitions in Magnetic Systems}

\author{Anthony HAMS and Hans De RAEDT}

\address{Institute for Theoretical Physics and
Materials Science Centre,\\
University of Groningen, Nijenborgh 4,
NL-9747 AG Groningen, The Netherlands}

\author{Seiji MIYASHITA and Keiji SAITO}

\address{
Department of Applied Physics, School of Engineering \\ 
University of Tokyo, Bunkyo-ku, Tokyo 113}

\date{\today}
\maketitle
\begin{abstract}
We examine the effect of the dynamics of the internal magnetic field
on the staircase magnetization curves observed in large-spin molecular magnets. 
We show that the size of the magnetization steps depends sensitively on 
the intermolecular interactions, even if these are very small compared to the
intra-molecular couplings.
\end{abstract}

Magnetization dynamics of nanoscale magnets, i.e. systems like Mn$_{12}$-acetate
and Fe$_{8}$ have been studied experimentally and theoretically 
lately~\cite{exp1,exp2,exp3,exp5,exp6,wern1999,peren1998}.
At sufficiently low temperatures quantum effects are observed, due to the discreteness
of the energy levels involved. 
When the magnetization of a crystal of such molecules is measured during a 
sweep of the external magnetic field, a staircase hysteresis loop is obtained. The steep parts of the 
staircase
correspond to the values of the external magnetic field where there is a crossing
of adiabatic energy levels. 
Several aspects of this quantum effect were studied in 
\cite{miya95,miya96,RMSGG97,MSD98,Dobro97,Gunther97}.
In a zero-temperature calculation, one finds that the magnetization can only
change in steps, very similar to the steps observed in recent experiments on
high-spin molecules Mn$_{12}$-acetate and Fe$_{8}$. At every crossing, only two levels play
a role and the transition probability can be calculated using
the
Landau-Zener-St\"{u}ckelberg (LZS) mechanism  \cite{Landau,Zener,St}.
Two parameters determine the LZS transition:
The energy 
splitting at the crossing and the sweep rate of the magnetic field.

The size of the energy splitting which leads to a  LZS transition probability is determined
by the off-diagonal terms in the Hamiltonian describing the system.
A straightforward perturbative calculation shows
that this splitting is roughly scales like
$\Gamma^{2|\Delta m|}$ where $\Gamma$
determines the magnitude of the off-diagonal terms and $\Delta m$
denotes the difference in magnetisation of the two relevant levels.
In the absence of a transverse applied field
the energy-level splittings in the high-spin molecules mentioned above
are so small that the probability for a single LZS transition is
effectively zero, unless the applied longitudinal field is rather large
(see for example ~\cite{peren1998}).

In the crystal the magnetic field felt by a particular molecule is the sum of the external field
and the internal field due to the presence of other magnetic molecules. As the inter-molecular
magnetic couplings in these materials are weak compared to the intra-molecular interaction between 
the spins, it seems reasonable to consider the former as a perturbation.
The purpose of this paper is to demonstrate that this argument
fails in the case of LZS transitions.
The point is that the LZS transition probability
depends on the rate of change of the effective magnetic field
{\em at the crossing}, which  can be changed significantly by the 
presence of the internal magnetic field.
The magnetization steps are found to be strongly affected by the type of interactions
among molecules.
We call this mechanism Feedback Effect on  Magnetization Steps (FEMS).

We first illustrate the effect for the case of the Mn$_{12}$-acetate molecules.
As a model Hamiltonian for this $S=10$ system we take~\cite{peren1998}
\begin{eqnarray} \label{eq:perenboom}
{\cal H} &=& -D_1\,S_z^2-D_4\,(S_x^4+S_y^4+S_z^4) 
\nonumber \\ & &
- c\,t\,\sin \theta \,S_x
- (c\,t\,\cos \theta+\lambda\,\langle S_z \rangle) \,S_z.
\end{eqnarray}
Compared to the model of~\cite{peren1998}
the extra feature in Hamiltonian~(\ref{eq:perenboom}) is the presence of a
mean-field term, the strength of which we parametrize by $\lambda$.
It is clear that in this mean-field approach any new effect
appears as a result of global changes of the internal field
generated by all the molecules and is not due to local fluctuations
which should be treated separately~\cite{stamp}.

\begin{figure}
\begin{center}
\epsfxsize=8.5cm 
\epsffile{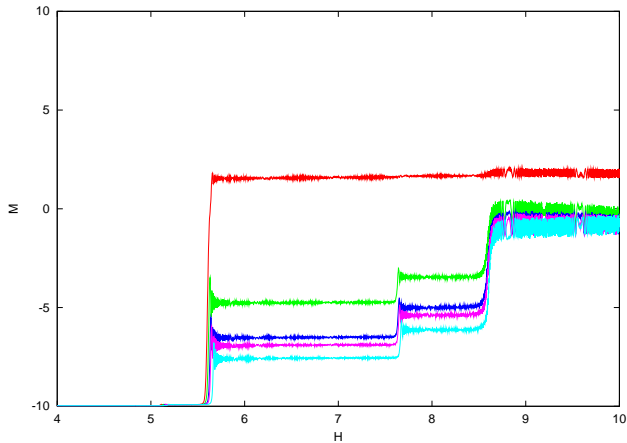}
\end{center}
\caption{Magnetization dynamics of the 
Mn$_{12}$-acetate model~(\ref{eq:perenboom}), for several values 
for the intermolecular coupling, 
$\lambda = -0.005$ (top curve), 
$-0.003$, 
$-0.001$, 
$0$, 
$0.003$ 
(bottom curve).
} 
\label{fig:mn12sweep}
\end{figure}

Quantitative results for the zero-temperature non-equilibrium dynamics
of model~(\ref{eq:perenboom}) can only be obtained through
a numerical integration of the Schr\"odinger equation.
Using standard techniques~\cite{RMSGG97} we compute the
magnetization steps for several values
of $\lambda$. The results
for $D_1 = 0.64$, $D_4 = 0.004$, tilt angle $\theta = 1^{\circ}$
and sweep rate
$c = 0.001$ (see~\cite{peren1998}, we use dimensionless units throughout
this paper)
are shown in Fig.~\ref{fig:mn12sweep}.
It is clear that the dynamics of the internal field can change the size of the magnetization steps considerably.
FEMS is observed for all $\lambda \neq 0$.
Note that the values of $|\lambda|$ we used
are not unrealistic ($|\lambda| \approx D_4 \ll D_1$), 
but rather small if we relate $\lambda$ to the dipole-dipole interaction which would yield a 
$\lambda$ which is $10-100$ times larger\cite{wernpriv}.

At very low temperatures experiments~\cite{peren1998} show
steps at lower values of $H$ than the ones at which we observe steps
in our calculation.
In fact, for the set of model parameters
given in~(\ref{eq:perenboom})
a much slower sweep rate $c$, much too slow for numerical calculations,
is required if we want to study the effect of
the internal field at all level crossings.

Therefore it is expedient to turn to a toy model inspired by
the one used to describe Fe$_{8}$~\cite{wern1999}.
We take a $S=2$ model with the following Hamiltonian~\cite{wern1999}:
\begin{eqnarray} \label{eq:wern}
    H &=& -D\,S_z^2 + E (\,S_x^2-\,S_y^2) \nonumber \\
&&+ \Gamma\,S_x
        - (c\,t+\lambda\,\langle S_z \rangle) S_z,
\end{eqnarray}
where we take $D = 1$, $E = 0.08$ and $\Gamma = 0.08$.
These parameters are chosen
such that we get two steps with a probability of about one half.

\begin{figure}
\begin{center}
\epsfxsize=8.5cm 
\epsffile{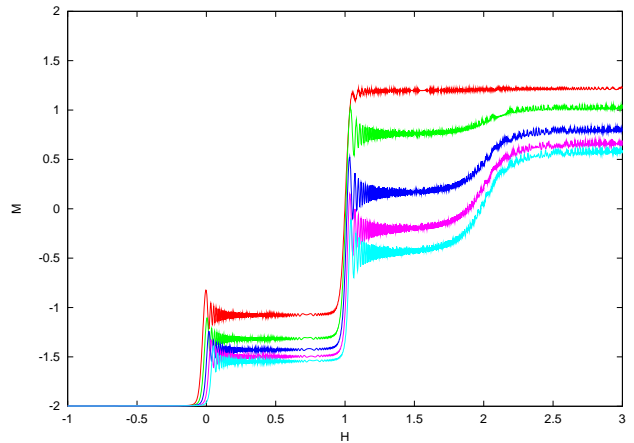}
\end{center}
\caption{Magnetization dynamics of the  $S=2$ model~(\ref{eq:wern}), for several values 
for the intermolecular coupling, 
$\lambda = -0.03$ (top curve), $-0.02$, $-0.01$, $0$, $0.01$ (bottom curve).}
\label{fig:spin2sweep}
\end{figure}

In Fig.~\ref{fig:spin2sweep} we show  the magnetization during a sweep of the magnetic field,
with a sweep rate $c = 0.01$, for several values of $\lambda$.
We see that the FEMS effect is large.
The transition probabilities are given in Table~\ref{tab:lambdadep}.
We clearly see a large change in the transition probabilities
due to the presence of the internal field.
\begin{table} 
\begin{tabular}{|l|r|r|r|r|r|}
$\lambda$ & $-0.03$ & $-0.02$ & $-0.01$ & $0$ & $0.01$ \\ \hline
Step 1 & 0.23 & 0.17 & 0.15 & 0.13 & 0.12  \\
Step 2 & 0.90 & 0.78 & 0.59 & 0.48 & 0.40 
\end{tabular}
\caption{Transition probabilities corresponding to the steps in 
Fig.~\ref{fig:spin2sweep}.}   
\label{tab:lambdadep}
\end{table}

A deeper understanding of the origin of the FEMS effect can be obtained
by considering the system of $N$ $S=1/2$ molecules
described by the Hamiltonian
\begin{equation}
{\cal H} = \sum^N_{i=1}\left[-\Gamma\,\sigma^x_i-J\,\sum^N_{j > i} 
\sigma^{z}_i\,\sigma^{z}_j+\,c\,t\,\,\sigma^z_i \right],
\end{equation}
where $c$ is the sweep rate, $\Gamma$ is the transverse field and $J$ determines the interaction 
strength between the molecules ($|J| \ll \Gamma$).
For simplicity we consider couplings between $z$-components only and assume the coupling
between the molecules to be the same. 
Since $|J|$ is small, we assume that we can make a mean-field-like 
approximation. The occurrence of FEMS does not depend on these simplifications (see below).
This yields
 a Hamiltonian of a single molecule in a background field:
 \begin{equation} \label{eq:hamil}
{\cal H} = -\Gamma\,\sigma_x-(c\,t+\lambda\,\langle \sigma_z \rangle)\,\sigma_z,
 \end{equation}
where $\lambda \propto J$ is an effective interaction.
The system is prepared in the ground state, corresponding to a large negative time $t$,
and the magnetic field is swept with constant velocity, until a large positive time is reached.
Then, in the LZS case with $\lambda=0$, the transition probability is 
given by the well-known LZS
formula
$p= 1-\exp(-\pi\,\Gamma^2/c)$.
For $\lambda \neq 0$  we write the Schr\"odinger equation corresponding to~(\ref{eq:hamil}) in component form:
\begin{eqnarray} 
\label{eq:start1}
i\,u' &=& \left( -c\,t-\lambda\,(2\,|u|^2-1)\right) \,u-\Gamma\,d,\\
i\,d' &=& \left( c\,t+\lambda\,(2\,|u|^2-1)\right) \,d-\Gamma\,u,
\label{eq:start2}
\end{eqnarray}
where we also have the normalization condition $|u|^2+|d|^2 = 1$.
From numerical simulations we (see below) find that the
tunneling is suppressed (enhanced) by the presence of a feedback term with 
positive (negative) $\lambda$.
This can be understood in terms of a changed
effective sweep rate at the point of the transition. Because the effective magnetic field at the 
position of the molecule is given by $c\,t+\lambda\,(2\,|u|^2-1)$, the effective sweep rate
would be $c+\lambda\,d\,\langle \sigma_z \rangle/dt$.

\begin{figure}
\begin{center}
\epsfxsize=8.5cm 
\epsffile{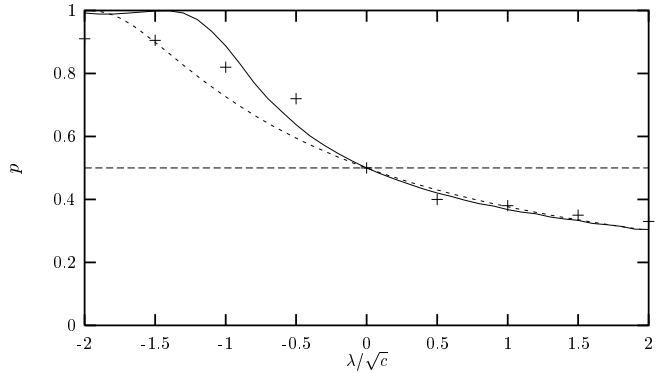}
\end{center}
\caption{Transition probability as a function of $\lambda/\sqrt{c}$, with $\Gamma^2/c$ such that
in the LZS case, $p=1/2$. The solid line is based on a numerical 
integration of~(\ref{eq:hamil}), the crosses are taken from a simulation of four interacting
$S=1/2$ spins
and the dashed line is based on~(\ref{eq:pguess}).}
\label{fig:pguess}
\end{figure}

If $\lambda$ is small but non-zero, the mean field term only contributes at the 
point of the crossing.
So we look at a Taylor expansion around the point of the transition $t_c$ (to be determined later),
\begin{equation}
u(t) = u_0 + u_1\,(t-t_c) +\order ((t-t_c)^2),
\end{equation}
and a similar expression for $d(t)$.
We insert this expansion in~(\ref{eq:start1}) and ~(\ref{eq:start2}) and obtain
\begin{eqnarray}
i\,\tilde u'(\tilde t) &=& -\tilde c\,\tilde t\,\tilde u(\tilde t)-\Gamma\,\tilde d(\tilde t),\\
i\,\tilde d'(\tilde t) &=& \tilde c\,\tilde t\,\,\tilde d(\tilde t)-\Gamma\,\tilde u(\tilde t),
\end{eqnarray}
with $\tilde t = t -t_c$, $\tilde c = c + 4\,\lambda\,\Re (u_0\,u_1^{*})$ is the renormalized sweep rate,
and were $\Re(u)$ is denoting the real part of $u$. 
We define $t_c$ as the point at which $c\,t +\lambda\,\langle \sigma_z \rangle$
changes sign, so
\begin{equation}
t_c = \frac{\lambda}{c}\left( 1-2\,|u_0|^2\right). 
\end{equation}
This enables us to write
$\tilde u_0 = u_0$ and $\tilde u_1 = u_1$.
To determine these constants we use 
Zener's solution~\cite{Zener,Abramowitz} and the properties of Weber 
functions. We find
\begin{equation}
|u_0|^2 = \frac{\pi}{4}\,\delta\,e^{-\pi \,\delta/4} \left| 
  \frac{1}
    {\Gamma(1 + \frac{i\,\delta}{4})} \right|^2
=
\frac{1}{2}\left( 1-(1-p)^2\right) ,
\end{equation}
where 
$p$ is the new probability for crossing, 
i.e. $p = 1-\exp(-\pi\,\Gamma^2/\tilde c)$ and $\delta = \Gamma^2/{\tilde c}$, and
$t_c = \lambda\,(1-p)^2\,/c$. The shift of the field at which the transition occurs
can be written as $\Delta H = \lambda\,(1-p)^2$.
To determine $\tilde c$ we calculate
\begin{equation} \label{eq:reexact}
\Re (u_0\,u_1^{*}) 
=
\sqrt{\tilde c}\,\frac{\pi\,\delta}{2} e^{-\delta\,\pi/4}
\Re 
\frac{e^{ -i\,\pi/4}}{\Gamma\left( 1 +\frac{i\,\delta}{4}\right) \,\Gamma\left( \frac{1}{2} -\frac{i\,\delta}{4}\right) }
. 
\end{equation}
We find that~(\ref{eq:reexact}) can be approximated by
\begin{equation} \label{eq:exact}
\Re (u_0\,u_1^{*}) 
\approx
\sqrt{\frac{\pi\,\tilde c}{8}}\, \delta\, e^{-\delta\,\pi/4}, 
\end{equation}
with an error of maximally $10\%$ (see Fig.~\ref{fig:pguess}).
Within this approximation, $\tilde c$ is given by the implicit equation 
\begin{equation} \label{eq:gammaapprox}
\tilde c \approx c + \lambda\,\Gamma^2
\sqrt{2\,\pi/\tilde c}\, e^{-\Gamma^2\,\pi/4\,\tilde c} .
\end{equation}
A simple relation can be obtained by replacing $\tilde c$ by $c$ on the right hand side. Then
\begin{equation} \label{eq:firstorder}
\tilde c \approx c + \lambda\,\Gamma^2\,
\sqrt{2\,\pi/c}\, e^{-\pi\,\Gamma^2/4\,c} ,
\end{equation}
and
\begin{equation}
\label{eq:pguess}
p \approx 1-\exp\left( -\frac{\pi\,\Gamma^2}{c}\frac{1}{1 + \lambda\,\Gamma^2\,
\sqrt{2\,\pi/c^3}\, e^{-\Gamma^2\,\pi/4\,c}}\right).
\end{equation}
The resulting probabilities are shown in Fig.~\ref{fig:pguess}.
The resulting probabilities based on a numerical solution
of~(\ref{eq:exact}) or~(\ref{eq:gammaapprox})
for  $\tilde c$ show similar behavior.
Also shown are the results obtained from the
exact numerical solution of the Schr\"odinger equation~(\ref{eq:hamil}).
As a test of the validity 
of the mean-field approximation we also show the result of four interacting $S=1/2$ spins, where
we assumed $\lambda = (N-1)\,J$.
Clearly the exact results confirm the validity of
the mean-field approximation
and
the simple analytic expression~(\ref{eq:pguess}).

For values of $\lambda$ below approximately $-2.0$ the description in terms
of a renormalized sweep rate breaks down, which can also be seen from the singularity in the argument of the 
exponent in eq.~(\ref{eq:pguess}). This is because the picture of a simple, 
single crossing breaks 
down and the effective magnetic field at the position of the spin will no longer be a strictly increasing 
function of time.
We conclude that the expression~(\ref{eq:pguess}) captures the main features of FEMS at a single crossing.

The relevant parameters, controlling the size of FEMS, are
$\Gamma/\sqrt{c}$ and $\lambda/\sqrt{c}$.
Only for S=1/2 the energy-level splitting is directly proportional
to $\Gamma^2$. For the high-spin molecules this is not case (see above),
in particular for the levels with large $|m|$.
Although in these cases the effective energy level-splitting
that enters the approximate two-level description can be small,
a rather small value of $\lambda$ can nevertheless change
the transition probability significantly.

We have shown that the magnetization steps in
the hysteresis loops of clusters of high-spin molecules may depend
sensitively on the change of the internal magnetic field at these steps.
This implies that the dynamics of this internal fields
has to be incorporated in a description of the magnetization dynamics,
even if its magnitude appears to be small compared
to 
the other model parameters (for large spin).
At finite temperatures the effect described in this paper
will be enhanced further due to the thermalization 
to states with lower energy and larger magnetization~\cite{Saito99}.

We would like to thank W.~Wernsdorfer for illuminating discussions.
 Support from the Dutch ``Stichting Nationale Computer Faciliteiten (NCF)'' and from
the Grant-in-Aid for Research of the Japanese Ministry of Education, Science and Culture
is gratefully acknowledged.

\end{document}